\documentclass{emulateapj}
\usepackage{psfig}

\begin{document}

 \received{}
 \accepted{}


\title{ The Ages of Globular Clusters in NGC~4365 Revisited with Deep HST Observations\altaffilmark{1}}

\author{Arunav Kundu\altaffilmark{2}, Stephen E. Zepf\altaffilmark{2}, Maren Hempel\altaffilmark{2}, David Morton\altaffilmark{2,3}, Keith M. Ashman\altaffilmark{3}, Thomas J. Maccarone\altaffilmark{4}, Markus Kissler-Patig\altaffilmark{5}, Thomas H. Puzia\altaffilmark{6}, Enrico Vesperini\altaffilmark{7}}

\altaffiltext{1} {Based on observations made with the NASA/ESA Hubble Space Telescope, obtained at the Space Telescope Science Institute, which is operated by    the Association of Universities for Research in Astronomy, Inc., under NASA contract  NAS 5-26555.}
\altaffiltext{2}{Michigan State University, East Lansing, MI 48824-2320, e-mail: akundu@pa.msu.edu, zepf@pa.msu.edu, hempel@pa.msu.edu}
\altaffiltext{3}{Univ. of Missouri, Kansas City, MO 64110, e-mail: dlmnn7@umkc.edu, ashman@umkc.edu}
\altaffiltext{4}{Univ. of Southampton, Highfield, U.K. SO17 1BJ, email: tjm@astro.soton.ac.uk}
\altaffiltext{5}{European Southern Observatory, 85748 Garching, Germany, e-mail: mkissler@eso.org}
\altaffiltext{6}{Space Telescope Science Institute, Baltimore, MD 21218, email: tpuzia@stsci.edu}
\altaffiltext{7}{Drexel University, Philadelphia, PA 19104, e-mail:vesperin@physics.drexel.edu}

\begin{abstract}

	We present new Hubble Space Telescope (HST)-NIC3, near-infrared H-band photometry of globular clusters (GC) around NGC~4365 and NGC~1399 in combination with archival HST-WCPC2 and ACS optical data. We find that NGC~4365 has a number of globular clusters with bluer optical colors than expected for their red optical to near-infrared colors and an old age. The only known way to explain these colors is with a significant
population of intermediate-age (2-8 Gyr) clusters in
this elliptical galaxy. In contrast, NGC~1399 reveals no such population. Our result for NGC~1399 is in agreement with previous spectroscopic work that suggests that its clusters have a large metallicity spread and are nearly all old. In the literature, there are various
results from spectroscopic studies of modest samples of NGC~4365
globular clusters. The spectroscopic data allow for either the presence or absence of a significant population of intermediate-age clusters, given the index uncertainties indicated by comparing objects in common between these studies and the 
few spectroscopic candidates with optical to near-IR colors indicative of intermediate ages. Our new near-IR data of the NGC~4365 GC system with much higher signal-to-noise agrees well with earlier published photometry and both give
strong evidence of a significant intermediate-age
component. The agreement between the photometric and spectroscopic results for NGC~1399 and other systems lends further confidence to this conclusion, and to the effectiveness of the near-IR technique.  
\end{abstract}


\section{Introduction}

 Globular clusters (GC) are invaluable probes of the major star formation episodes in the life of a galaxy because each individual GC has a specific age and metallicity which reflects the physical conditions at the epoch of its formation. Thus, the observed color and spectrum are
much easier to interpret than the complex superposed populations seen in 
integrated light (e.g. Ashman \& Zepf 1998).  Despite the simple nature of the stellar population of a GC, determining both the age and metallicity of any unresolved population requires overcoming the well known age-metallicity degeneracy that causes both increasing age and increasing metal content to have similar effects on optical colors and spectral features.

 One way to break this age-metallicity degeneracy is to combine optical and near-infrared colors, as the near-IR is mostly sensitive to the metallicity of the giant branch while optical colors are affected by both metallicity and age.
Puzia et al. (2002) (hereafter P02) employed this technique on ground-based, VLT K-band observations in combination with WFPC2 data to study two early type galaxies, NGC~3115 and NGC~4365. P02 found that the globular cluster system of NGC~4365 has a significant intermediate-age (2-8 Gyrs old) component, which has no counterpart in the predominantly old NGC~3115 clusters. The discovery of intermediate age GCs in a fairly typical elliptical such as NGC~4365 is a powerful illustration of the ability of GCs to probe major formation episodes of galaxies. 
Subsequent spectroscopy of a handful of bright GCs in these galaxies (Larsen et al. 2003, hereafter L03; Kuntschner et al. 2002), and studies of the cluster systems of several other galaxies with both optical to near-IR photometry (Hempel et al. 2003) and spectroscopy (Puzia et al. 2005) agree on the age and metallicity distribution determined by the two techniques. However, a recent spectroscopic analysis of a small sample of NGC~4365 clusters by Brodie et al. (2005) (hereafter B05) 
suggests that the previously photometrically and spectroscopically identified intermediate age GCs are instead an old population. Given  the interest in determining whether there are intermediate age GCs in early type galaxies and its implications on how galaxies form, it seems important to analyze independent data for NGC~4365.

 We have obtained deep H-band images of NGC~4365 using the NIC3 camera on board the HST in order to study the mass function of its GCs. We present a new, entirely HST-based study of the cluster system of NGC~4365. We compare our analysis to the aforementioned published studies, and a control sample in NGC~1399 using the exact same HST instruments, to comment on the constraints on the age distribution of the intermediate metallicity GC population.

\section {Observations}

 We obtained deep, dithered, H-band (F160W), NICMOS-NIC3 observations of 5568s each at three positions in NGC~4365 on 17$^{th}$ Nov, 2003, 15$^{th}$ Jun, 2004 and 17$^{th}$ Jun 2004 for our HST program GO-9878. These observations coincided with the WF chips of archival WFPC2 V (F555W, 2200s) and  I (F814W, 2300s) images obtained on 31$^{st}$ May 1996. The galaxy was observed in the g (F475W, 750s), and z  (F850LP, 1120s) with the ACS on 6$^{th}$ Jun, 2003. NGC~1399 was imaged in the H (F160W, 384s) with the NIC3 on 18$^{th}$ Dec, 1997, the B (F450W, 5200s) and I  (F814W, 1800s) with the WFPC2 on 2$^{nd}$ Jun, 1996, and the g (F475W, 760s) and z  (F850LP, 1130s) on 11$^{th}$ Sep, 2004 with the ACS.

The NGC~4365 NIC3 observations are sub-pixel dithered at four positions. Using the drizzle algorithm (Fruchter \& Hook 2002) we reconstructed high resolution images, alleviating the effect of the undersampled 0.2'' NIC3 pixels. Importantly, the dithering placed the center of each GC at different locations with respect to the center of a pixel, thus reducing the effects of intrapixel sensitivity variations in the NIC3 array (Xu \& Mobasher 2003). The NICMOS observations of NGC~1399 were obtained with NIC1 and NIC2 in focus. Since NIC3 does not share a common focus with these instruments the NIC3 images are out of focus. However, the instruments on the HST are well studied and characterized; hence we were able to extract valuable information from these data. We both drizzled the NGC~1399 NIC3 images and shifted and added them on the original scale. The photometry determined from each image was in excellent agreement. We chose to use the shifted images to minimize possible uncertainties due to centering issues in out of focus images. Interestingly, the out of focus nature of the NGC~1399 NIC3 image mitigates intrapixel sensitivity effects. The WFPC2 data for both galaxies were dithered by integer pixels. The images were shifted and added to remove cosmic rays and charge traps. After inspection we also analyzed the drizzled, geometric distortion corrected ACS images of both galaxies.

	Candidate GCs were identified in the WFPC2 and ACS images using the constant S/N detection technique described in Kundu et al. (1999).  The lists, and the results described below for each set are in good agreement. We use the ACS selected candidates due to the higher S/N, and the slightly improved ability to distinguish between GCs and contaminating objects.

 Aperture photometry was performed in each image using zeropoints from the HST data handbook. Aperture corrections from small radii were measured from our data to account for the partially resolved GC profiles in HST images (Kundu et al. 1999). Foreground reddening corrections from Schlegel, Finkbeiner, \& Davis (1998) were also applied.  {\it All} photometry in this paper is reported in the Vegamag system. Where applicable we have transformed ABMAG magnitudes to Vegamag using zeropoint offsets from Sirianni et al. (2005).

While a 0.5$''$ aperture was used for photometry in the NGC~4365 NIC3 images, the NIC3 observations of NGC~1399 did not have a small core so a large 5 pixel radius (1$''$) aperture was used for aperture photometry. Aperture corrections were determined from TinyTim models (Krist 1995). Although the NICMOS focus
history indicated that the Pupil Alignment Mechanism (PAM) position for best focus for the NGC 1399 NIC3 images  should be near -13mm, comparison of point sources with TinyTim models  revealed that a PAM position of -10mm provided the best fit to the data. Varying the center of an object within a pixel and allowing PAM positions between -5mm and -16mm changed the correction by 0.04 mag rms. This factor was added in quadrature to the photometric uncertainty. 

\section {Results \& Discussion}

 In the rest of this analysis we study the GCs in the g-I, I-H plane because both galaxies have been observed in these filters with the same instruments, and this choice of filters provides the largest baseline for both optical and infrared colors. The conclusions of this study are unaffected by the choice of optical and infrared color baselines selected from the filter set available to us.

	Figure 1 plots the g-I vs I-H colors for 70 GCs in NGC~4365 and 11 GCs in NGC~1399 with photometric uncertainties less than 0.1 mag in each color. The least luminous source in the g, I and H filters plotted in Fig 1 are 25.83$\pm$0.07, 23.40$\pm$0.05, and 21.43$\pm$0.07 in NGC 4365 and 23.62$\pm$0.02, 21.74$\pm$0.02, 20.08$\pm$0.10 in NGC 1399 respectively. Fiducial lines of constant metallicity and constant age from the simple stellar population models of Bruzual \& Charlot (2003) (hereafter BC03) are also plotted. It is apparent that there is a significant excess of GCs with blue g-I colors for a given I-H color in NGC~4365 as compared to NGC~1399. Such colors can only be explained by the presence of an intermediate age population of GCs younger than $\approx$8 Gyrs. In contrast the GCs in NGC~1399 appear to be primarily old. 

The relative paucity of GCs in NGC~1399 is because it represents a single out of focus NIC3 field. Although the lower S/N of the NGC~1399 NIC3 image causes a preferential selection of red, young and/or metal-rich GCs, most of the metal-rich GCs in NGC~1399 appear to be older than 10 Gyr despite this bias. This suggests that the overwhelming majority of GCs in this galaxy are old with a handful of possible young ones. This is completely consistent with the spectroscopic analyses of GCs in NGC~1399 by Kissler-Patig et al. (1998) and Forbes et al. (2001) who found that the majority of their samples of 18 and 10 GCs respectively are old, with a range of metallicities extending to roughly solar. In consonance with our results, each of these studies identified two possible/likely young GCs in their respective samples.

	The color-color plot of NGC~4365 GCs provides a striking contrast to NGC~1399 with a clear excess of GCs younger than $\sim$8 Gyrs, consistent with the conclusions of P02 and the spectroscopic follow up by L03.  Figure~2 compares our data with  Anders \& Fritze-v. Alvensleben (2003) and Maraston (2005) models. While the models differ in detail due to the choice of stellar track and calibration technique, the  conclusion that NGC~4365 has a large population of intermediate age GCs and NGC~1399 does not is independent of the choice of model.

\subsection{Statistical Significance of the Age Distributions}
The direct determination of the epochs and efficiency of the major episodes of star formation from color-color plots are complicated by issues like selection effects and photometric errors. We choose instead to apply the modelling technique of Hempel \& Kissler-Patig (2004). In brief, we create input models with two populations of clusters using Monte Carlo simulations and stellar models. The cumulative age distribution of a range of input models is then compared to the data to find the distribution that best fits the data.

Figure 3 plots the cumulative fraction of GCs that are older than a given age in each of the galaxies, based on BC03 models. This analysis is restricted to GCs with $[Fe/H]\gtrsim-0.4 dex$ to minimize the effects of incompleteness. Figure 3 shows that the NGC~1399 GCs are older on average, with a median age of $\approx$12 Gyr as compared to $\approx$5 Gyr for NGC~4365. 
Next we fixed the age of the older population to 13 Gyr and conducted the simulations described in Hempel \& Kissler-Patig (2004) for a two burst model. Figure 4, which plots the reduced $\chi^{2}$ comparing the model with the data, shows that approximately 60\% of the NGC~4365 metal-rich GCs in our field of view  were formed about 4 Gyrs ago. The corresponding fraction for NGC~1399 is only about 20\%, although given the selection biases in NGC 1399 this is likely an upper limit on the constraints. We note that only a handful of old, metal-poor GCs are observed in NGC 4365. This is not surprising since the optical color magnitude diagram (Kundu \& Whitmore 2001) suggests that GCs with colors of V-I$\approx$0.95 corresponding to the typical metal-poor peak are fainter than the other GCs in NGC 4365, as is expected for older clusters. We shall investigate the luminosity and mass functions of NGC 4365 GCs in a future paper. 

\subsection{Comparison with Previous Observations}

As discussed above, the inferred age and metallicity distributions of our program galaxies are in good agreement with previous photometric and spectroscopic studies (P02, L03, Kissler-Patig et al.~1998, Forbes et al.~2001). The recent spectroscopic analysis of NGC~4365 by B05 however reaches a different conclusion suggesting  that the intermediate age GCs are actually old, intermediate metallicity clusters. Unfortunately, the B05 study has very few GCs with known near-IR and optical colors.

Specifically, of the large sample of NGC~4365 GCs with optical and near-IR colors there are 12 shown in Fig. 8 of B05 as having spectroscopic data. Of these, 9 have been observed by L03 providing reasonable overlap with the photometric data. B05 observed 3 GCs with known near-IR and optical colors, including two in L03. The remaining two of the 12 are an object with a given position more than 2$''$ from any candidate in P02 and thus an uncertain photometric counterpart, and one with unreliable colors due to likely cosmic ray events in the WFPC2 image. B05 observed a total of 6 objects with intermediate optical colors; the three above with near-IR colors, and three without such colors to allow an age estimate (see Fig 1). A recent shallow near-IR imaging study by Larsen et al. (2005) observed additional B05 clusters and suggested a broad spread of GC colors consistent with P02 and this study. However, the authors express reservations about their own calibration. An important aspect of the data presented here are the much smaller error bars which both clearly show that GC colors indicate an age spread and are consistent with the shallower data sets.

In Figure 5, we plot the g-I, I-H colors of the six matches with the L03 data and the subset of two sources observed by B05 that fall within our fields of view. We note that most of these GCs appear to be of intermediate age with moderate metallicity, in agreement with the results of L03. Given the small overlap with the B05 observations, model uncertainties, and one candidate that lies on the 8 Gyr isochrone and is consistent with an old age, no strong conclusions can be drawn from the B05 data. We also note that B05 found significantly different Lick index values compared to L03 for the three GCs in common between the two studies.
The up to $3\sigma$ difference between the indices of these objects in the age sensitive Balmer lines indicates that the uncertainties in these index measurements, and therefore the error in the age and metallicity constraints in one or both spectroscopic studies are underestimated. On the other hand, the photometric analyses of the NGC~4365 GCs agree unambigously. Given the agreement of these and L03 for NGC~4365, and the agreement between the NGC~1399 spectroscopic and photometric results, we believe the preponderance of evidence points towards the presence of intermediate age GCs in NGC~4365. To determine the full spatial extent of this intermediate-age population will require deep near-IR imaging over a wider field.

\section{Conclusions}

Globular clusters provide fossil records of the  major star formation episodes in a galaxy. Thus, constraining the ages and metallicities of cluster sub-populations provides important insight into the process of galaxy formation and evolution. The near-IR and optical color photometric technique and spectroscopy  measure the ages and metallicities of clusters by probing different physical phenomena and provide independent sanity checks. We have analyzed the cluster system in the inner regions of NGC~1399 and NGC~4365 and conclude that the metal-rich clusters in NGC~1399 are predominantly old while many of the corresponding clusters in NGC~4365 are of intermediate age. These results are in good agreement with previous spectroscopic and photometric studies of these galaxies and hence gives us further confidence in the photometric technique.

This research was supported by STScI grant HST-GO-09878.01-A and NASA-LTSA grants NAG5-11319 and NAG5-12975.

\begin{figure*}[!ht]
\includegraphics[angle=-90, scale=0.6]{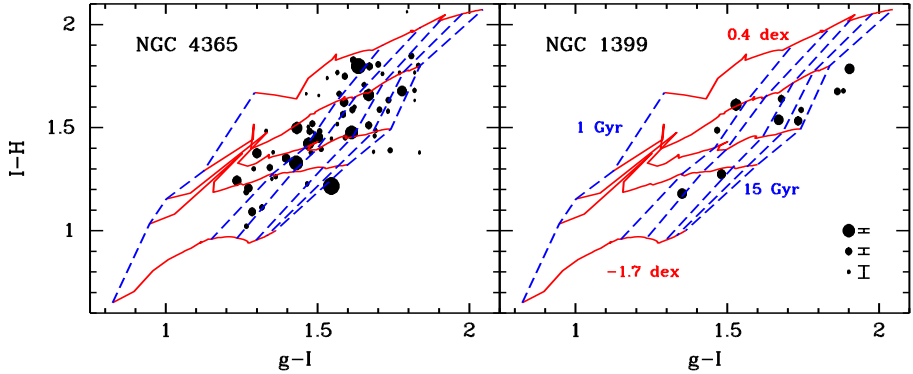}
\caption{g-I vs I-H plots of globular clusters in NGC~4365 and NGC~1399. Only GCs  with uncertainties less than 0.1 mag in each axis are shown. The lines trace age (1, 3, 5, 8, 11 and 14 Gyr from the left) and metallicity ([Fe/H]=-1.7, -0.7, -0.4, 0 and 0.4 dex from the bottom) contours from BC03 models. The size of the points are inversely proportional to the I-H uncertainty as shown at the bottom right. The g-I uncertainties are comparable to the I-H values in NGC 4365, and much smaller than the I-H estimates in NGC 1399. }
\end{figure*}

\begin{figure*}[!ht]
 \includegraphics[angle=-90, scale=0.6]{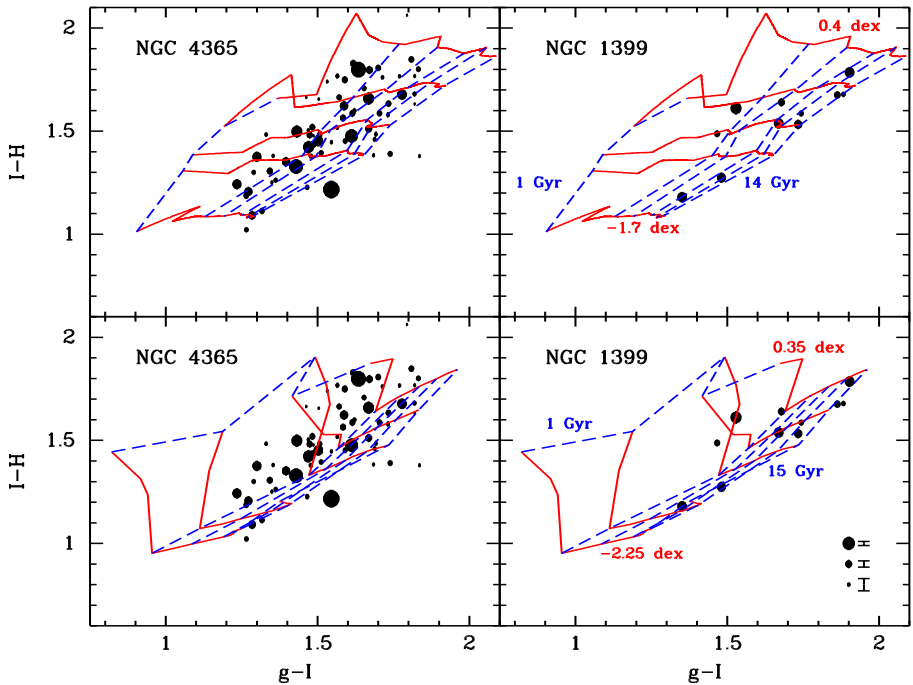}
 \caption{ As in Fig 1, with Anders \& Fritze-v. Alvensleben (2003) models  in the upper panels and Maraston (2005) models in the lower ones. Lines in the upper plots represent [Fe/H]=-1.7, -0.7, -0.4, 0  and 0.4 from the bottom and 1, 3, 5, 8, 11 and 15 Gyrs from the left.  Lines in the lower panels trace [Fe/H] -2.25, -1.35, -0.33, 0 and 0.35 dex  from the bottom and 1, 3, 5, 8, 11 and 15 Gyrs  from the left.   }
 \end{figure*}

\begin{figure}
\epsscale{0.6}
\plotone{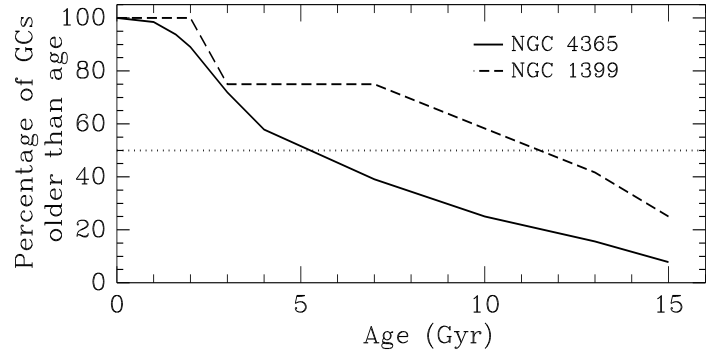}
\caption{ Cumulative fractional age distributions of the GCs in NGC~1399 and NGC~4365 from BC03 models.  }
\end{figure}

\begin{figure}[!ht]
\epsscale{0.4}
\plotone{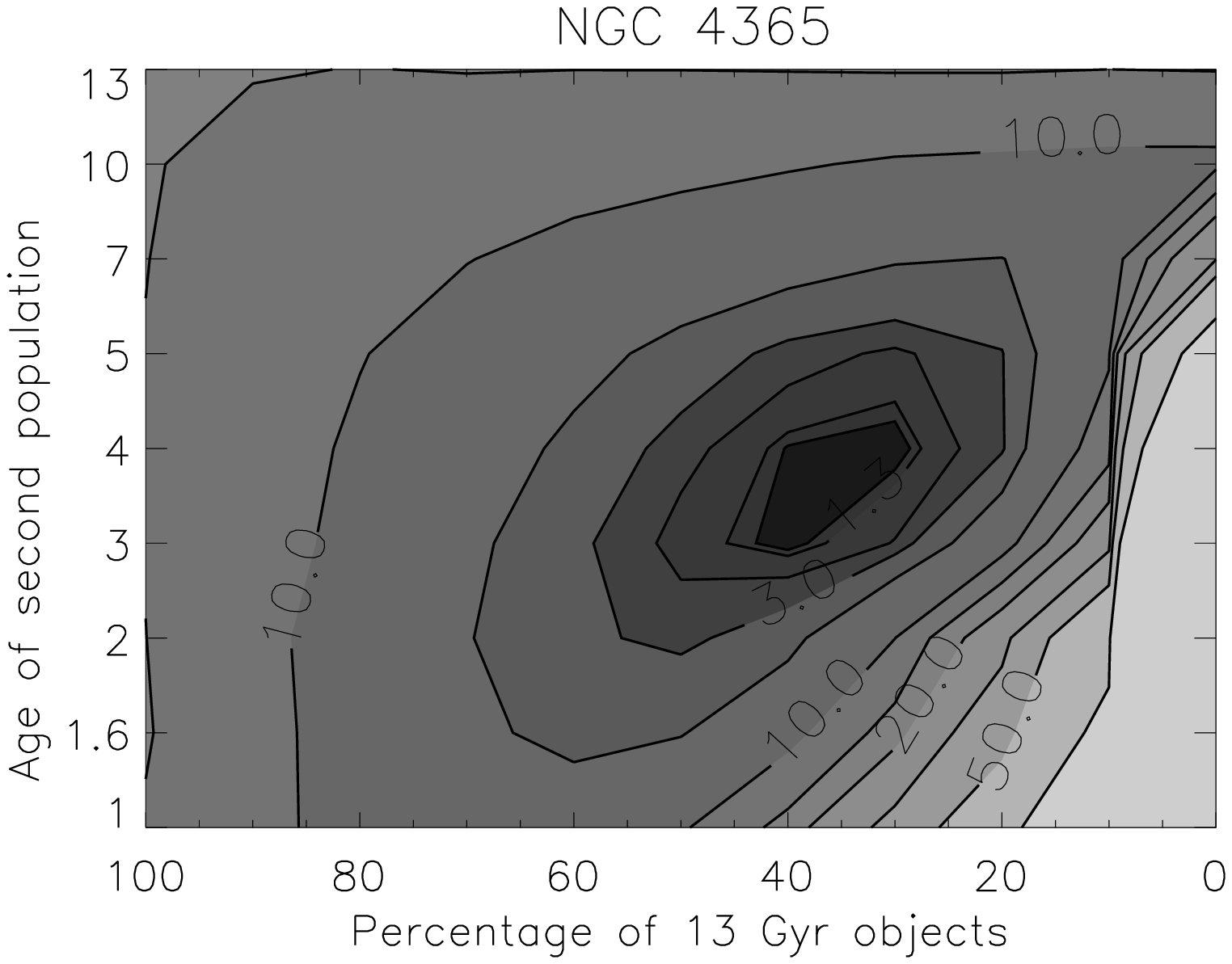}\\
\plotone{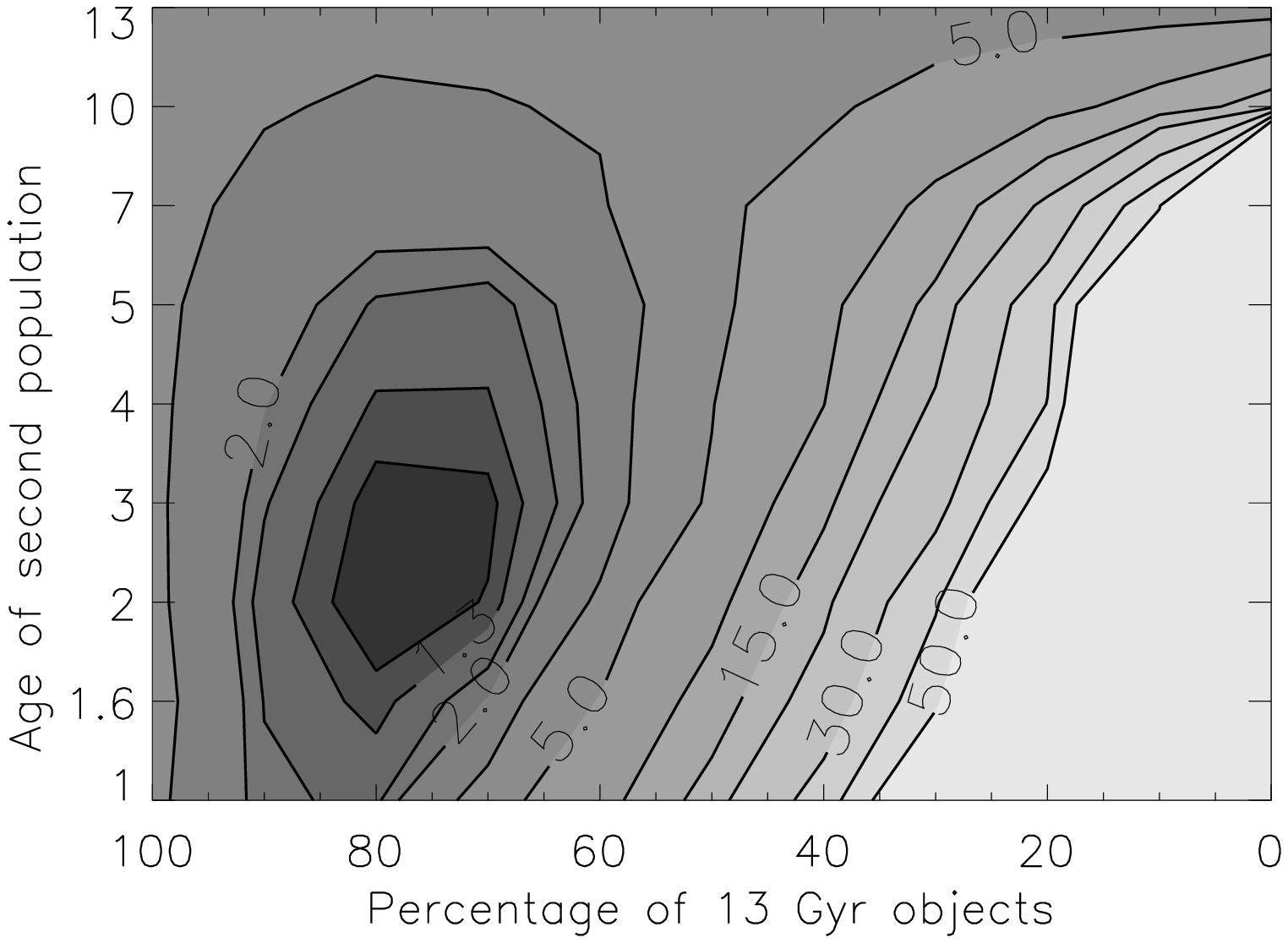}
\caption{ Reduced $\chi^{2}$ plots comparing the age distributions of GCs in NGC~4365 and NGC~1399  with simulated cluster systems.}
\end{figure}

\begin{figure}[!ht]
\includegraphics[angle=-90, scale=0.6]{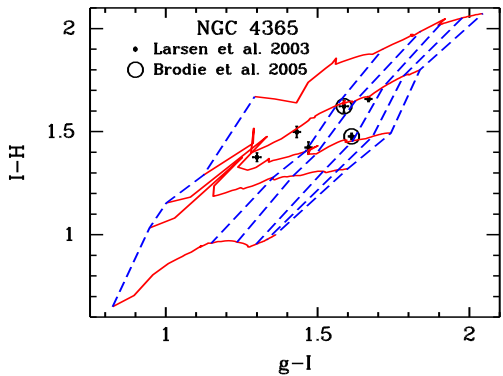}
\caption{ g-I vs I-H plots of GCs studied spectroscopically. Fiducial lines trace BC03 models as in Fig 1. }
\end{figure}

\end{document}